\documentclass[sigconf,screen]{acmart}

\AtBeginDocument{%
  \providecommand\BibTeX{{%
    \normalfont B\kern-0.5em{\scshape i\kern-0.25em b}\kern-0.8em\TeX}}}

\setcopyright{acmlicensed}
\copyrightyear{2018}
\acmYear{2018}
\acmDOI{XXXXXXX.XXXXXXX}

\acmConference[ESEM '24]{18th International Symposium on Empirical Software Engineering and Measurement}{Oct 24--25,
  2024}{Barcelona, Spain}

\acmISBN{978-1-4503-XXXX-X/18/06}

\usepackage{graphicx} 
\usepackage{url}
\usepackage{booktabs}
\usepackage{multirow}
\usepackage{tcolorbox}
\usepackage{subcaption}

\begin{document}
\title{Documenting Ethical Considerations in Open Source AI Models}
\author{Haoyu Gao}
\affiliation{
 \institution{The University of Melbourne}
  \country{Victoria, Australia}
 }
\email{haoyug1@student.unimelb.edu.au}

\author{Mansooreh Zahedi}
\affiliation{
\institution{The University of Melbourne}
\country{Victoria, Australia}
}
\email{mansooreh.zahedi@unimelb.edu.au}

\author{Christoph Treude}
\affiliation{
\institution{Singapore Management University}
\country{Singapore}
}
\email{ctreude@smu.edu.sg}

\author{Sarita Rosenstock}
\affiliation{
\institution{The University of Melbourne}
\country{Victoria, Australia}
}
\email{sarita.rosenstock@unimelb.edu.au}

\author{Marc Cheong}
\affiliation{
\institution{The University of Melbourne}
\country{Victoria, Australia}
}
\email{marc.cheong@unimelb.edu.au}


\begin{abstract}
    \textbf{Background:} 
    The development of AI-enabled software heavily depends on AI model documentation, such as model cards, due to different domain expertise between software engineers and model developers. From an ethical standpoint, AI model documentation conveys critical information on ethical considerations along with mitigation strategies for downstream developers to ensure the delivery of ethically compliant software. However, knowledge on such documentation practice remains scarce.
    \textbf{Aims:} The objective of our study is to investigate how developers document ethical aspects of open source AI models in practice, aiming at providing recommendations for future documentation endeavours.  
    \textbf{Method:} We selected three sources of documentation on GitHub and Hugging Face, and developed a keyword set to identify ethics-related documents systematically. After filtering an initial set of 2,347 documents, we identified 265 relevant ones
    and performed thematic analysis to derive the themes of ethical considerations. 
    \textbf{Results:} Six themes emerge, with the three largest ones being model behavioural risks, model use cases, and model risk mitigation. 
    \textbf{Conclusions:} Our findings reveal that open source AI model documentation focuses on articulating ethical problem statements and use case restrictions. 
    We further provide suggestions to various stakeholders for improving documentation practice regarding ethical considerations.
\end{abstract}



\maketitle

\section{Introduction}

AI-enabled software refers to software systems containing AI algorithm components which facilitate learning and problem solving~\cite{ozkaya2020really}. It is used in a range of systems, including critical domains such as health care~\cite{zhou2018unet++}, law enforcement~\cite{raaijmakers2019artificial}, and education~\cite{ciolacu2018education}, due to rapid advances in AI techniques in recent  years~\cite{rombach2022high, radford2021learning, touvron2023llama}. However, biases and instances of dangerous model use are still frequently encountered, posing potential harm to end users. For example, investigations have revealed cases where opaque AI-enabled software predicts a higher risk of future crime for African-Americans compared to white individuals~\cite{propublicaMachineBias}.  Therefore, ensuring adherence to ethical best-practice, vis-à-vis the behaviour of AI models, becomes important. Recent pushes for fairness, transparency, and accountability in AI systems~\cite{facctconferenceFAccTStatement} give rise to proposals to incorporate these concerns in documentation practices for software developers. The most prominent proposals, including Mitchell et al.'s model cards~\cite{mitchell2019model} and Gebru et al.'s datasheets~\cite{gebru2021datasheets} have been eyed by regulators as potential loci of AI governance paradigms. In particular, model card has become the default introductory document in Hugging Face registries~\cite{hfcard}.

From a software engineering perspective, software documentation encompasses essential information regarding system design, architecture, and the development process, serving as a valuable means of communication among stakeholders~\cite{gao2023evaluating}. The completeness and usability of software documentation are two of the critical factors relevant to developers~\cite{aghajani2020software}. However, documentation is found to be one of the collaboration challenges in machine learning (ML) enabled systems, as team communication frequently relies on verbal exchanges lacking clear documentation~\cite{nahar2022collaboration}. Documenting ethical considerations is particularly important, as it informs software developers about potential ethical risks of the underlying models and the corresponding mitigation methods. Bhat et al.~\cite{bhat2023aspirations} found that ethics-related discussions are shallow in current documentation practice.
However, there is limited empirical evidence on what ethical considerations are documented for the current AI models. By understanding the current practice in AI ethics documentation, we can identify specific documentation issues, and provide guidelines for both documentation maintainers and policy makers. This paper intends to bridge this gap, and provide concrete suggestions for aiding the documentation process.

We conducted an empirical study of the ethics-related AI model documentation practice on two major open source platforms, GitHub and Hugging Face, specifically focusing on both model cards and README files. We refer to these two types of artefacts as AI model documents consistently throughout the paper. We collected an initial set of 2,347 documents, and developed a comprehensive keyword set to identify ethics-related documentation. After filtering the non-related documents and the ones that have identical ethical documentation, we performed thematic analysis on 265 documents. Our analysis shows that documentation about ethics in the model cards revolves around six themes including: (1) Data quality concerns, (2) Model behavioural risks, (3) Model risk mitigation, (4) Model use cases, (5) Reference to other materials, (6) Others. We found that most of the documentation is about model behavioural risk, while data quality concerns and suggestions for ethical risk mitigation are less discussed. In addition, we identify that most ethics-related risks discussed are only briefly outlined, and the already scarce risk mitigation suggestions are not concrete and actionable. We discuss the implications for various stakeholders and provide concrete suggestions. The replication package is at \url{https://zenodo.org/records/12531181}.

The key contributions of our research are as follows: (1) a keyword set that effectively captures documentation on ethics-related issues in open source documents, (2) a comprehensive set of themes on ethical considerations in the current AI model documents as found in our dataset, and (3) a replication package with the dataset and our qualitative analysis to facilitate future research.




\section{Related Work}

\subsection{Ethics in Software Engineering}


A pragmatic framework for discussing ethics is \textit{principlism}, which encompasses principles such as respect for autonomy, nonmaleficence, beneficence, and justice~\cite{cheong2023ethical}. Schwartz recognised ten value categories including
security and conformity, encompassing 58 human values~\cite{schwartz2005basic, schwartz2006basic}. The values, such as helpfulness, are used in mapping with ethical considerations in software engineering. Some of the values are incorporated in the software process management frameworks and decision making~\cite{ferrario2014software, ferrario2016values} or to measure values in software engineering~\cite{winter2018measuring}. Using this framework, Alidoosti et al.~\cite{alidoosti2022ethics} conducted a systematic literature review on software engineering ethics, discovering relationships among different ethical values, namely value conflicts and value tensions. They identified stakeholders that are affected by ethical considerations in software engineering, namely that system users, system development organisations, and indirect stakeholders are mostly concerned with safety. In addition, Perera et al.~\cite{perera2020study} conducted a study investigating publications in top software engineering venues pertaining to various human values. They found that values such as helpfulness, privacy, and environmental protection are discussed in software engineering research. Several studies, both empirical and algorithmic in nature, were proposed to enhance helpfulness~\cite{steinmacher2016overcoming}, privacy~\cite{dijk2017model}, and environmental protection~\cite{morales2018earmo}.


Bias and fairness are crucial aspects of ethics in modern software development~\cite{chakraborty2021bias}. Brun and Meliou~\cite{brun2018software} visioned fairness challenges across software engineering phases, including requirement specification, architecture design, testing, and verification. Tools are needed to address root causes of bias, such as issues in requirements, algorithms, implementation, or data. Galhotra et al.~\cite{galhotra2017fairness} proposed a testing-based method to measure software discrimination. In the domain of machine learning, various techniques are proposed. The bias in machine learning models can be mitigated with pre-processing, in-processing, and post-processing techniques~\cite{mehrabi2021survey}. For pre-processing, techniques are used to transform the data so that the underlying discrimination is removed~\cite{d2017conscientious}. The in-processing techniques aim to incorporate changes into the objective function or impose constraints during model training~\cite{berk2017convex}. For post-processing, Gohar et al.~\cite{gohar2023towards} identified ensemble as a way to mitigate bias in machine learning models. To complement existing work on ethical aspects in software engineering and to support the developers of AI-enabled software, in this study, we investigate the documentation practice for ethical considerations in AI models.

\subsection{Machine Learning Documentation}
Machine learning documentation reports information about the model and dataset, aiming to provide transparency among stakeholders about the systems~\cite{mitchell2019model}. Regarding data, datasheets proposed by Gebru et al.~\cite{gebru2021datasheets} suggest documenting content such as motivation, composition, collection process, and recommended uses. Pushkarna et al.~\cite{pushkarna2022data} proposed an at-scale framework of Data Cards for better transparency and explainability, with information across a dataset's lifecycle. Hutchinson et al.~\cite{hutchinson2021towards} proposed a framework to make visible the often overlooked work and decisions involved in dataset creation, covering dataset requirements, design, implementation testing, and maintenance.

For model documents, Model Cards~\cite{mitchell2019model} and FactSheets~\cite{arnold2019factsheets} are proposed standard frameworks for AI models. In particular, Model Cards have received substantial interest. They serve as the default document in Hugging Face, one of the largest open source machine learning model platforms, and are adopted by companies such as Google and Salesforce for their public models~\cite{hfcard}. The proposed framework in Model Cards includes model details, intended use, factors, metrics, evaluation data, training data, quantitative analyses, ethical considerations, and caveats and recommendations.

Despite the proposal of machine learning documentation frameworks, document practice effort and quality still fall short. A recent study interviewed 45 practitioners from 28 organisations and identified documentation as one of the biggest challenges when building and deploying machine learning systems into production~\cite{nahar2022collaboration}. One of the participants in their study stated ``team members often collect information and keep track of unwritten details in their heads''. Therefore, similar to traditional software document issues~\cite{aghajani2019software}, completeness and usability issues, among others, also appear in machine learning documentation.

Recent studies have started examining machine learning documentation practices to provide insights for developing more comprehensive guidelines or documentation tools. Bhat et al.~\cite{bhat2023aspirations} quantitatively studied the model card documentation practice alignment with the proposed framework, and identified that most documentation has an unbalanced coverage of aspects in the model card proposal. Yang et al.~\cite{yang2024navigating} quantitatively investigated Hugging Face dataset documentation, shedding light on current dataset documentation practices. Pepe et al.~\cite{pepe2024hugging} studied the transparency perspective of pre-trained transformer models dataset, bias, and license declaration documented in Hugging Face model cards. However, none of the previous work focused on ethics. Our work contributes to this field by examining machine learning documentation practices in ethics-related content. Our results inform developers of AI-enabled software by providing a comprehensive view of how ethical considerations are documented.

\section{Methodology and Study Design}
In this section, we introduce our research methodology and the design of our study.

\subsection{Research Questions}

We structured our investigation into two research questions (RQs).

\textbf{RQ1}:\textit{ How accurately can we detect ethics-related discussions within different types of documentation?}  This research question focuses on the development of a set of keywords to detect ethical content within AI model documentation. We employed expansion techniques and experimented with the removal of keywords that resulted in false positives to derive the final set of keywords.

\textbf{RQ2}:\textit{What is documented about ethics in AI-enabled open source software projects?} This research question focuses on identifying the ethical considerations documented within open source projects targeted at the developers of AI-enabled software. We conducted thematic analysis to address this question.




\subsection{Data Collection and Preprocessing}
We selected two Open Source Software (OSS) platforms as data source, namely GitHub and Hugging Face. GitHub is the largest OSS platform, hosting various types of software projects, while Hugging Face is the largest model registry, enabling machine learning model reuse. In terms of the artefact, we focused on ``model cards''. However, there is no standardised location for model card documents in OSS. For example, \texttt{aparrish/pincelate} provides its model card within the README file rather than having a dedicated location.  Only searching for separate artefacts would miss such cases. Therefore, README files are also included as our target.

For GitHub, we used the search API to find repositories containing an explicit \texttt{model\_card.md} or \texttt{model-card.md} document, as well as the ones mentioning the term ``model card'' in their README files. We distinguish these two data sources, and refer them as to \texttt{GH\_CARD} and \texttt{GH\_README}. For \texttt{GH\_CARD}, we obtained 589 repositories in total. However, the GitHub search API has a limitation: for data sources exceeding 1,000 results, it cannot return the entire population~\cite{ghsearch}. Therefore, we only obtained a sample of 1,020 repositories for the \texttt{GH\_README} data source after one search. We further applied filters for the repositories, requiring them to have at least ten stars and no forks to ensure the quality of repositories selected~\cite{dabic2021sampling}, narrowing down the selection to 173 and 305, respectively for \texttt{GH\_CARD} and \texttt{GH\_README}. For Hugging Face, we systematically collected models that were among the top 2,000 downloaded in December 2023, and kept the ones having model card documents, ending up with 1,869 projects. We refer to the Hugging Face data source as \texttt{HF\_CARD}.

However, it is possible that certain documents within these repositories do not mention ethical considerations, as we did not initially filter for such content. Training machine learning models to identify whether documents are related to ethics requires a large amount of annotated data. Annotating such data would be time-consuming and require domain expertise. Therefore, we adopted a keyword-based filter to heuristically obtain the ethics-related documents.

\begin{figure*}[t]
    \centering
    \includegraphics[width=1.6\columnwidth]{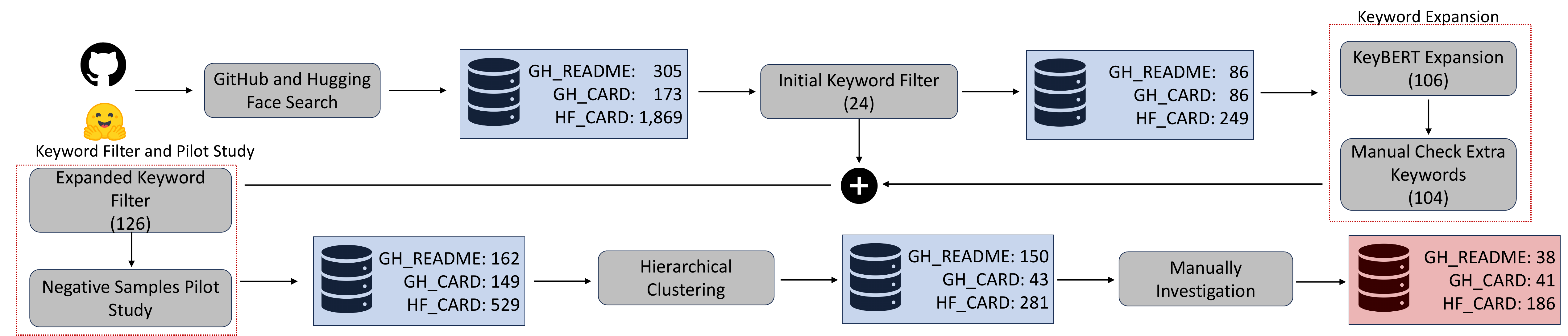}
    \Description[]{}
    \caption{Data Collection Steps}
    \label{fig:collection-process}
\end{figure*}

Given the absence of a gold standard for ethical keywords in mining documents, we adopted an iterative keyword expansion method akin to prior work~\cite{shahin2023study}. Figure~\ref{fig:collection-process} describes our filtering steps. Specifically, we initially selected a set of keywords, including ``ethics'', terms in ``FAccT'' (fairness, accountability, and transparency), plus "morality" and "responsibility," as identified in a systematic literature review~\cite{pant2022ethics}. Then, we searched the noun, verb, adjective, and adverb forms of those words using a tool from GitHub~\cite{ghword}. Forms that deviate from their ``ethical'' meaning would lead to false positives (e.g., the adverb variant "fairly" of the adjective "fair"), and we did not include them within our base keyword set.

After filtering with the base keywords, we obtained relevant \texttt{GH\_CARD}, \texttt{GH\_READE}, and \texttt{HF\_CARD} documents, totaling of 86, 86, and 249, respectively. We manually filtered out false positive documents within this set, resulting in 86, 27, and 222 documents. False positives refer to documents that do not contain ethics-related content or repositories that are not machine learning registries or machine learning applications. In \texttt{GH\_CARD} and \texttt{HF\_CARD}, the main reason for false positives was keyword usage, such as "bias" being used as a statistical term. Meanwhile, most false positives occurring in \texttt{GH\_README} are because the repositories do not contain machine learning registries or applications.

Based on the selected documents, we expanded the keyword set using KeyBERT~\cite{keyBert}. The basic idea for KeyBERT is to extract the most relevant words from a given text based on the cosine similarity between each of the words and the whole text. As our documents cover diverse topics such as model description and performance, utilising the entire document for keyword extraction may yield suboptimal results. Meanwhile, topics are relatively stable within the granularity of paragraph. Consequently, we used paragraphs that contain at least one of our base keywords and applied KeyBERT for the keyword expansion. 205 new keywords were generated, and two authors separately annotated them as relevant or not to ethics.  An initial Cohen's Kappa score of 0.61 was reached, indicating a substantial level of agreement~\cite{mchugh2012interrater}. Disagreements were resolved after discussion. This step finally provides 44 new keywords and is expanded to 106 considering all derivatives.

         


After integrating expanded keywords, we can detect more ethics-related documents, with 149, 172, and 656 for sources of \texttt{GH\_CARD}, \texttt{GH\_README} and \texttt{HF\_CARD}. To validate the efficacy of our expanded keywords, we randomly sampled a 95\% confidence rate subset of 107, 119, and 243 documents, respectively. We manually annotated the documents as relevant or irrelevant to ethical considerations. 6, 73, and 88 documents were identified as irrelevant, corresponding to a false positive rate of 5.6\%, 61.3\%, and 36.2\%, respectively.

\begin{figure}
    \centering
    \includegraphics[width=0.6\columnwidth]{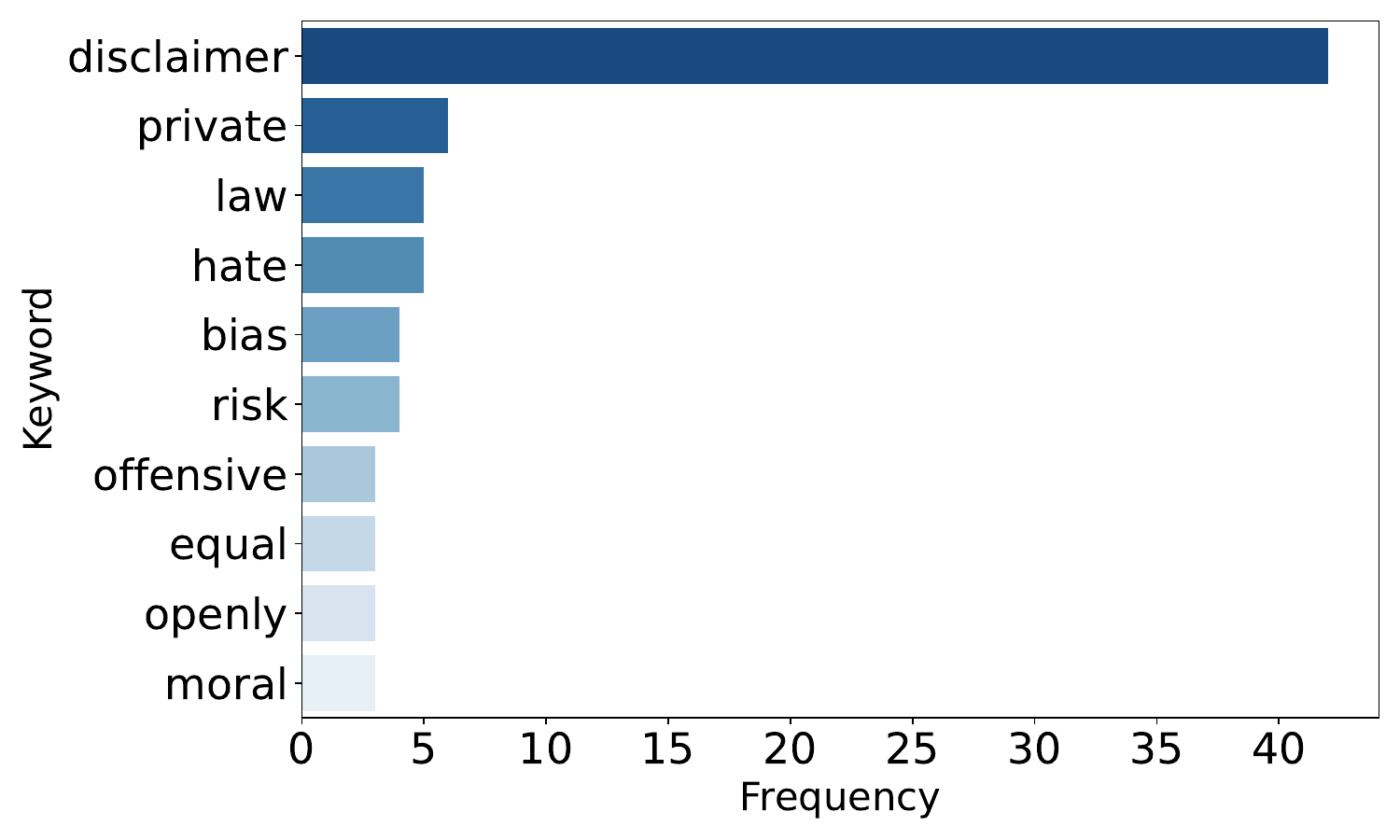}
    \caption{Top-10 False Positive Keywords}
    \label{fig:false-positive}
\end{figure}

Given that false positives in \texttt{GH\_README} are mostly due to repositories (e.g., non-software and non-ML repositories) rather than documents, we looked at keywords that cause false positives in the other two sources. As indicated in Figure~\ref{fig:false-positive}, the keywords ``disclaimer'' and ``private'' contribute to most of the false positives, while other keywords cause fewer errors. The keyword ``disclaimer'' is identified to appear in many Hugging Face model cards, stating that the documents were written by the Hugging Face team instead of the repository authors. Meanwhile, the keyword ``private'' is noisy, with false positive examples in ``private discussion channel'', and ``private dataset'' (i.e., not released). We set the frequency threshold at five and excluded two keywords exceeding it. Within our sampled dataset, these two keywords often occur by themselves when the data record is a false positive, and removing them would drop 46 false positives while only one true positive is ignored. Therefore, we removed them from our keyword set.

We further conducted another pilot study, randomly sampling 50, 21, and 50 negative documents from \texttt{GH\_CARD}, \texttt{GH\_README}, and \texttt{HF\_CARD} respectively. Only two repositories within \texttt{GH\_CARD} were supposed to be positive. This suggests that our chosen keywords effectively encompass the majority of ethics-related content, with minimal instances of oversight.

Therefore, to answer RQ1, we developed a comprehensive ethics-related keyword set to heuristically filter contents of AI model documentation. This filter can capture the vast majority of the ethics-related documents for model card documents. By applying these filters, we can immediately exclude documents that are clearly not-related to ethics. However, further manual efforts are still required to filter out false positives in the identified documents. The set of keywords is available in our replication package.

\subsection{Duplication Detection and Clustering}

In our previous manual investigation of ethics-related documents, we observed the phenomenon of document reuse across numerous repositories. Notably, many projects share highly similar, or even identical, contents with respect to ethical considerations. To mitigate the potential dominance of such repetitive documents in our analysis, we clustered the documents into distinct content groups. 

\begin{figure}[t!]
    \centering
    \includegraphics[width=0.6\columnwidth]{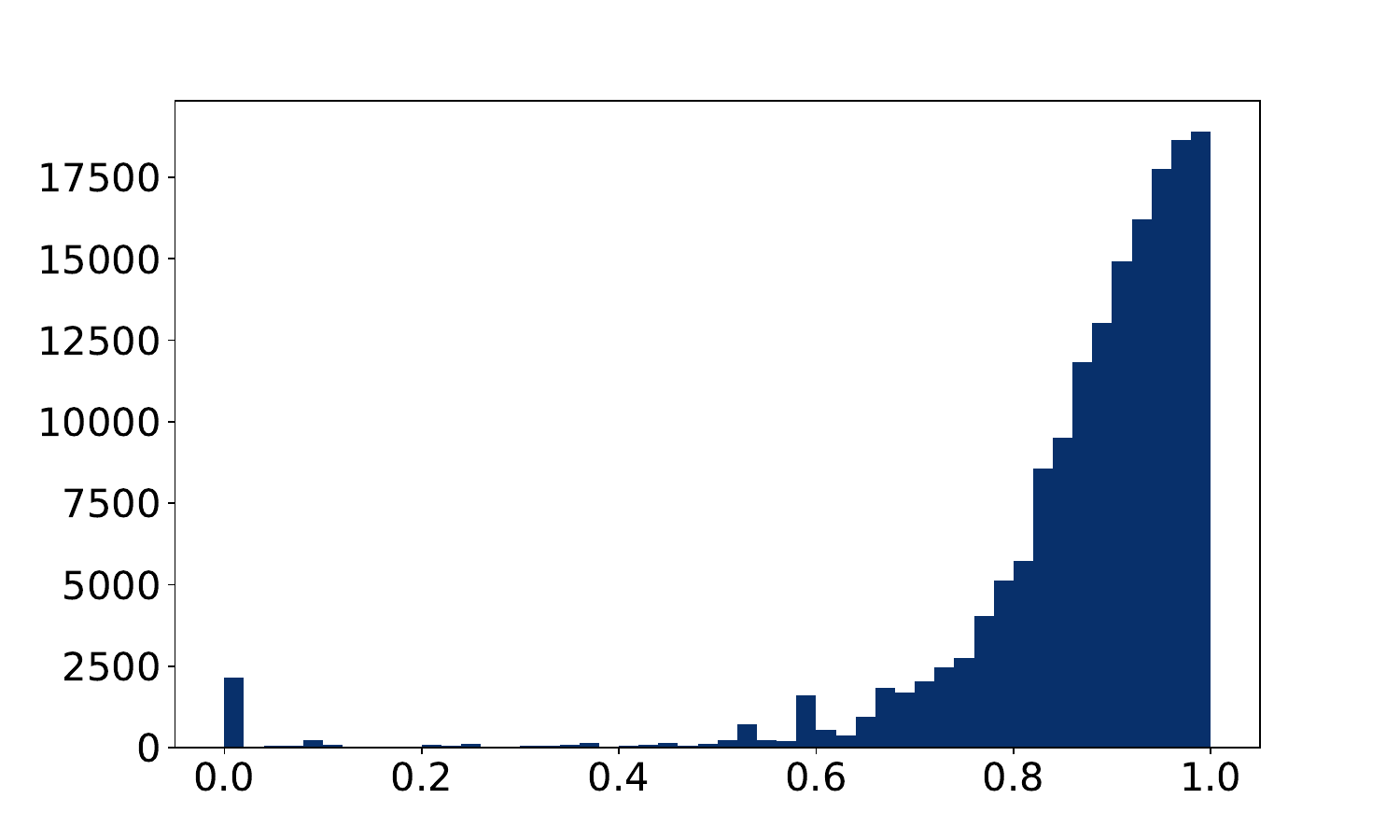}
    \caption{Pairwise Document Cosine Distance}
    \label{fig:distance-distribution}
    \Description[]{}{}
\end{figure}

In particular, for each document, we extracted paragraphs related to ethics using our curated keywords and calculated pairwise document distances using TF-IDF-based cosine distance. 
Figure~\ref{fig:distance-distribution} shows the distance distribution. A prominent peak is observed at 0, indicating identical contents, while a small number of documents exhibit similarity within the threshold of 0.1. Furthermore, the document reuse phenomenon primarily comprises instances of identical content, resulting in close distances between them. Meanwhile, most of the documents still have a larger distance. We used this distance matrix for agglomerative hierarchical clustering, truncating clusters at a distance threshold of 0.1. This yielded 43, 150, and 281 distinct clusters in \texttt{GH\_CARD}, \texttt{GH\_README}, and \texttt{HF\_CARD} respectively, indicating the ethical document reuse is more prevalent in model cards than in REDAME files. We randomly sampled one document from each cluster to conduct the following thematic analysis.

\begin{figure*}[t!]
    \centering
    \includegraphics[width=1.5\columnwidth]{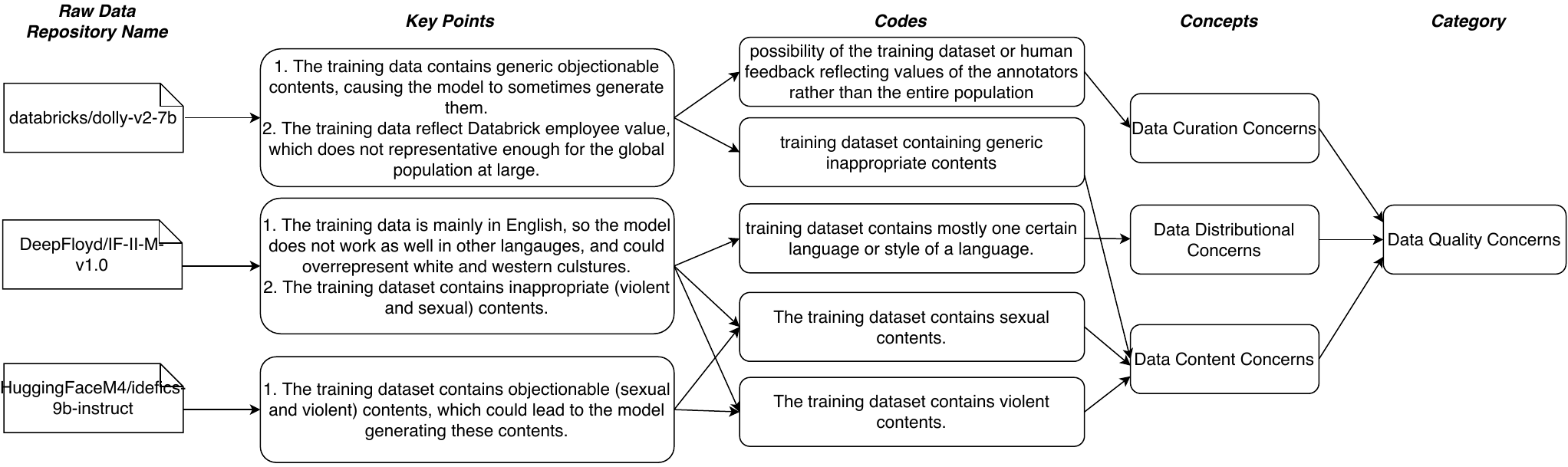}
    \caption{Steps for Coding Process}
    \label{fig:coding-process}
    \Description[]{}{}
\end{figure*}

\begin{figure*}[t!]
    \centering
    \includegraphics[width=1.5\columnwidth]{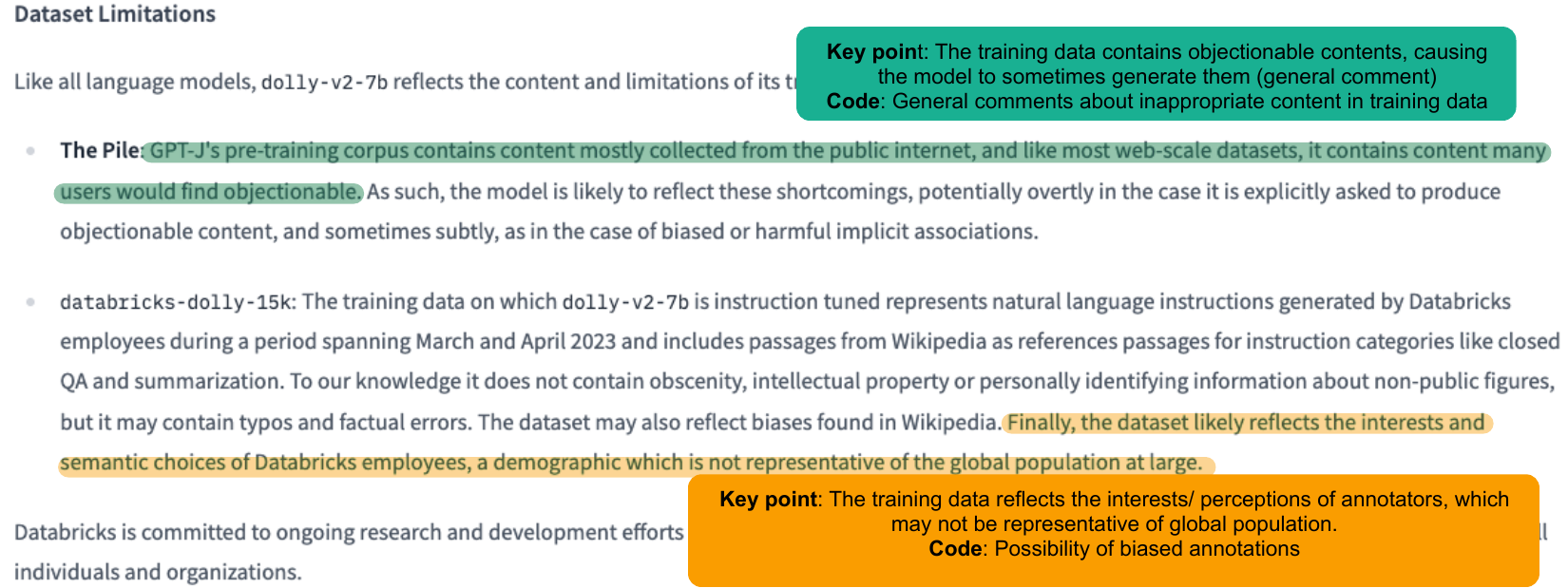}
    \caption{Coding Example for Raw Document (\texttt{databricks/dolly-v2.7b})}
    \label{fig:coding-example}
    \Description[]{}{}
\end{figure*}

\subsection{Data Analysis}

To cover the diverse range of ethics-related topics comprehensively without neglecting certain perspectives, four authors first looked at a sample of ten documents. We annotated the parts of the documents that contain ethical considerations independently, before discussing to reach agreement. After that, the first author analysed the remaining documents. Throughout the process, the first author flagged uncertain or ambiguous content for further discussion. Finally, 41, 38, and 186 documents from the source of \texttt{GH\_CARD}, \texttt{GH\_README} and \texttt{HF\_CARD} respectively, were included for the analysis. We used the following steps from thematic analysis~\cite{braun2006using, cruzes2011recommended} for the qualitative analysis stage of the included documents. 

\textit{Getting familiar with data}: The first author organised the documents and created spreadsheets for analysis, which are shared among all the authors. The first author extracted key points of the ethical discussions from the documents, as suggested to initiate the open coding~\cite{hoda2012self}. The second and third author closely examined this process. Disagreements were resolved during regular meetings.

\textit{Generating base level codes}: After agreeing on the extracted key points, the first author proceeded with open coding \cite{cruzes2011recommended}. During the process, careful investigation of key points along with raw data in the chosen documents were conducted to understand the context and ensure the accuracy of assigned codes. Throughout the process, regular discussions were held among the first three authors to verify the codes within and across different documents. This iterative process of code development led to continuous adjustment of the codes, with some being deleted, merged, or split.

\textit{Synthesising and finalising themes}: In this phase we grouped the emerging codes into higher level themes, as suggested in thematic analysis~\cite{braun2006using, cruzes2011recommended}. Here, we got inspiration from a template suggested by~\cite{mitchell2019model} for documenting model cards, when grouping the emerged codes from our data into themes. Finally, we generated a mind-map visualising all the themes and organisation of codes. The mind-map was used to facilitate discussion among all the authors and reaching to agreement. Figure~\ref{fig:coding-process} provides an illustrative example for generating one of the themes called ``Data quality concern'', and Figure~\ref{fig:coding-example} provides an example for coding on the raw data.



\section{Results}
We synthesised 81 codes belonging to six major themes, which are: (1) Data quality issues, (2) Model behavioural risks, (3) Model risks mitigation, (4) Model use cases, (5) Reference to other materials, and (6) Others. Table~\ref{tab:qualitative-result} presents the sub-themes under each theme, their frequencies, as well as the proportion for each theme covered in three different data sources. Since one document could cover multiple sub-themes, the ``Total'' is not the sum of the percentages in each sub-theme.  We also added the frequency for each sub-theme in parentheses after its description.

\begin{table*}[t]
    \caption{Qualitative Analysis Results}
    \centering
    \resizebox{0.68\textwidth}{!}{%
    \begin{tabular}{l l r r r r r r}
    \toprule
        Theme & Sub-theme & \# Codes & Frequency & Overall & \texttt{HF\_CARD} & \texttt{GH\_CARD} & \texttt{GH\_README} \\
        \midrule
        \multirow{4}{*}{Data quality concern} & Data curation concern & 2  & 5 & 1.89\%  & 1.08\% & 7.32\%  & 0.00\%  \\  
                                             & Data distributional concern & 6  & 23 & 8.68\% & 9.14\% & 12.20\% & 2.63\% \\ 
                                            & Data content concern  & 6  & 69  & 18.87\% & 19.89\% & 12.20\% & 21.05\% \\ 
                                            \cmidrule(l){2-8}
                                            & \textbf{Total} & \textbf{14} & \textbf{97} & \textbf{27.55\%} & \textbf{28.49\%} & \textbf{26.83\%} & \textbf{23.68\%} \\
        \midrule
        \multirow{4}{*}{Model behavioural risks} & Possibility of biased output. & 11 & 143 & 35.85\% & 39.25\% & 39.02\% &                                                 15.79\% \\
                                                 & Generation of, or reaction to objectionable content & 5 & 61 & 18.49\% & 22.04\% & 19.51\% & 0.00\% \\
                                                 & Correctness and reliability of model behaviour & 3 & 51 & 18.11\% & 20.97\% & 17.07\% & 5.26\% \\
                                                 \cmidrule(l){2-8}
                                                 & \textbf{Total} & \textbf{19} & \textbf{255} & \textbf{49.43\%} & \textbf{56.99\%} & \textbf{41.46\%} & \textbf{21.05\%} \\
        \midrule
        \multirow{4}{*}{Model risk mitigation} & Risk mitigation in the model development phase & 4 & 48 & 16.98\% & 18.28\% &                                                              19.51\% & 7.89\% \\
                                               & Risk mitigation in the post-model development phase & 4 & 13 & 4.91\% & 4.84\% & 7.32\% & 2.63\% \\
                                               & Risk mitigation suggestions for downstream developers & 6 & 108 & 28.30\% & 34.41\% & 17.07\% & 10.53\% \\
                                               \cmidrule(l){2-8}
                                               & \textbf{Total} & \textbf{14} & \textbf{169} & \textbf{43.02\%} & \textbf{48.39\%} & \textbf{41.46\%} & \textbf{18.42\%}\\
        \midrule
        \multirow{4}{*}{Model use cases} & Model term of use & 3 & 41 & 13.96\% & 17.20\% & 4.z87\% & 7.89\% \\
                                        & Out-of-scope use cases & 8 & 87 & 26.79\% & 28.49\% & 34.15\% & 10.53\%\\
                                        & Potential malicious use and misuse of the model & 8 & 82 & 21.13\% & 19.89\% & 26.83\% & 21.05\%\\
                                        \cmidrule(l){2-8}
                                        & \textbf{Total} & \textbf{19} & \textbf{210} & \textbf{48.68\%} & \textbf{50.53\%} & \textbf{56.10\%} & \textbf{31.58\%} \\
        \midrule
        \multirow{3}{*}{Reference to other materials} & Reference to in-house materials & 2 & 34 & 12.83\% & 12.37\% & 2.44\% & 26.32\% \\
                                                     & Reference to third-party external material & 5 & 60 & 21.51\% & 23.66\% & 19.51\% & 13.16\%\\
                                                     \cmidrule(l){2-8}
                                                     & \textbf{Total} & \textbf{7} & \textbf{94} & \textbf{31.70\%} & \textbf{32.26\%} & \textbf{21.95\%} & \textbf{39.47\%} \\

        \midrule

        \multirow{5}{*}{Others} & Limited exploration of ethical considerations & 3 & 33 & 12.45\% & 13.98\% & 12.20\% & 5.26\% \\
                                & Future socio-ethical research direction & 3 & 34 & 12.83\% & 12.90\% & 14.63\% & 10.52\% \\
                                & Model developers' disclaimer & 1 & 28 & 10.57\% & 11.83\% & 4.88\% & 10.53\% \\
                                & Environmental impact & 1 & 15 & 5.66\% & 4.84\% & 14.63\% & 0.00\% \\             
                                \cmidrule(l){2-8}
                                & \textbf{Total} & \textbf{8} & \textbf{110} & \textbf{39.25\%} & \textbf{40.86\%} & \textbf{43.90\%} & \textbf{26.32\%} \\
        
        \bottomrule
         
    \end{tabular}}
    \label{tab:qualitative-result}
\end{table*}

\subsection{Data Quality Concern}
The behaviour of machine learning models is significantly influenced by the quality of datasets used for their training~\cite{gong2023survey}. $\sim$27.55\% of the documents contain data quality concerns. We synthesised three sub-themes and discuss them as follows:

\textbf{Data content concern (69).}  When discussing data quality, "data content concern" is a frequently-occurring sub-theme. This relates to the presence of inappropriate content within certain entries in the dataset. Specifically, concerns revolve around the possibility of biased\footnote{Here, we use the nomenclature \textit{biased} to refer to discrimination and prejudice, from a socio-ethical lens.}, offensive, sexual, and violent content. For example, the model card of \texttt{DeepFloyd/IF-II-M-v1.0} notes, ``\textit{The model was trained on a subset of the large-scale dataset LAION-5B, which contains adult, violent and sexual content.}''. Additionally, private information issues were also raised. For example, \texttt{facebook/nllb-200-3.3B}  mentions, ``\textit{Although we invested heavily in data cleaning, personally identifiable information may not be entirely eliminated.}''. 

However, for seven of the documents, their discussion on data quality issues is vague (e.g., only stating ``\textit{low-quality data}''), and cannot be categorised specifically. Therefore, we refer to them as ``generic inappropriate content''.  We argue that these statements are not as useful, as they do not provide any specific, concrete directions for downstream developers to investigate.

\textbf{Data distributional concern (23).} As highlighted by Buolamwini et al.~\cite{buolamwini2018gender}, under-representative groups in the data can introduce bias into trained models. The data distributional concern refers to the discussion that certain data is statistically imbalanced in its distribution. The most mentioned distributional bias pertains to language and culture, as exemplified in the statement  ``\textit{... mostly trained on English text, other languages and cultures are insufficiently accounted for}" in \texttt{DeepFloyd/IF-II-M-v1.0}. Other distributional bias including style, gender, and data collection time, are also mentioned. Some discussions suggest internet data may be biased toward those with easier access. For example, \texttt{facebook/flava-full} notes that ``\textit{A larger portion of this dataset comes from internet and thus can have bias towards people most connected to internet such as those from developed countries and younger, male users (sic).}''

Similarly, we observed that five of the documents include more high-level comments about potential data distribution concerns (e.g., ``\textit{There may be distributional bias in the RedPajama dataset that can manifest in various forms in the downstream model deployment} in 
\texttt{cerebras/btlm-3b-8k-base}''), without providing more details. 

\textbf{Data curation concern (5).} The sub-theme of "Data curation concerns" is often overlooked. Unlike the previous two sub-themes, which primarily focus on the data, the way that data is curated reflects the interpretation and interests from human annotators. An example is how a ``\textit{...dataset likely reflects the interests and semantic choices of Databricks employees, a demographic which is not representative of the global population at large.}'', mentioned in the model card of project \texttt{databricks/dolly-v2-7b}. This caveats the thoroughness of ML developers when they consider potential ethical issues, due to factors of demography, subjectivity, and perspectives.

\begin{tcolorbox}[left=1pt, top=1pt, right=1pt, bottom=1pt]
    \textbf{Summary of data quality issue}: The main discussion related to ``data quality concerns'' focus on objectionable and private data. Data distribution concerns address under-represented groups, while data curation concerns involve biases from human annotators, often overlooked in ethical considerations in our dataset.
\end{tcolorbox}

\subsection{Model Behavioural Risks}
This theme encompasses various ethics-related risks associated with the behaviours of ML models. In our analysed data, this is the most frequently mentioned theme, with approximately half of the documents containing such discussions. We synthesised three sub-themes and discuss them as follows:

\textbf{Possibility of biased output (143).} This sub-theme relates to the phenomenon that certain groups perceive bias in the output of AI models. One of the key considerations in ethics is related to the biased behaviour of models. Among different factors, gender bias ($\sim21.6\%$) is the factor that is most frequently brought up, followed by culture and language ($\sim15\%$), racial ($\sim10\%$), occupation ($\sim5.8\%$), age ($\sim3.3\%$), religion ($\sim3.3\%$), and disability ($\sim1.7\%$). However, the largest proportion ($\sim38.3\%$) is related to generic mention of model output bias. These abstract mentions provide little helpful information to downstream software developers, and act more like a disclaimer from document writers' perspective.

Meanwhile, different documents vary in the level of detail for biased outputs. For instance, ``\textit{Based on known problems with NLP technology, potential relevant factors include bias (gender, profession, race and religion)}'' in \texttt{koboldAI/OPT-13B-Nerys-v2}, briefly mentions different bias factors without delving into the details. In contrast, \texttt{albert-base-v2}~\cite{hfalbert} provides code examples that demonstrate gender bias by showcasing the model's predictions for male and female inputs. Moreover, \texttt{martin-ha/toxic-comment-model}~\cite{hftoxic} includes a table detailing the model performance when considering various ethnic groups, genders, and disabilities. Offering comprehensive details on potential model biases can help downstream developers better understand the nature and quantify the severity of the issues.

Furthermore, there is also discussion on biased behaviours of models, stemming from an original pre-trained model, or being inherited after further fine-tuning. Document users can refer to the upstream models for additional model bias information.

\textbf{Generation of, or reaction to objectionable content (61).} In the era of Large Language Models (LLMs), machine learning models learn to generate responses to complex human requests. However, uncensored models may comply to unethical requests and generate objectionable content. 
Two frequently mentioned types of objectionable content are those that are rude and harmful to individuals ($\sim48.1\%$), and sexual content ($\sim23.1\%$). Notably, $\sim28.8\%$ of the mentioned entries only provide a generic warning that the generated content might be inappropriate, without further substantiation. For instance, \texttt{google/flan-t5-base} notes, ``\textit{As a result the model itself is potentially vulnerable to generating equivalently inappropriate content or replicating inherent biases in the underlying data.}''.  Such generic statements lack detailed insights that could be investigated by downstream developers.

Similarly, the levels of detail in documenting objectionable content vary. Notably, some documents provide model safety measurements on established benchmarks. For instance, one of the Llama model documents~\cite{hfllama} evaluates different versions of the model's toxic generation on a benchmark called "Toxigen". 

In addition to input, there are also mentions on the model lacking control over unethical requests. ``\textit{It will be highly compliant to any requests, even unethical ones.}''. These discussions provide directions for downstream developers to fix.

\textbf{Correctness and reliability of model behaviour (51).} The model correctness discussed here refers to the \textit{factual} accuracy of the output content, not performance accuracy. The term ``hallucination'' is often used, describing the phenomenon where the model produces self-contradictory or misaligned responses with established world knowledge~\cite{zhang2023siren}. In \texttt{bigscience/T0\_3B}, an example is provided where the model responds ``\textit{yes}'' to the question ``\textit{Is the earth flat?}'', indicating the model's agnosticity to real-world facts. However, most documents do not provide specific details, as we have observed and discussed earlier.

Other reliability factors are also discussed, albeit less frequently. These include the possibility of model output revealing information from training data and the lack of transparency in the model.

\begin{tcolorbox}[left=1pt, top=1pt, right=1pt, bottom=1pt]
    \textbf{Summary of model behavioural risks}: The most discussed sub-theme is potential biased model output. Additionally, model output can be objectionable or comply with unethical requests. Correctness and reliability issues focus on hallucinations, revealing training data, and lack of transparency.
\end{tcolorbox}

\subsection{Model Risk Mitigation}
Understanding what can be done to mitigate potential model risks is crucial, as mere awareness is not enough. This theme encompasses the efforts made by model developers as well as recommendations offered to downstream developers to mitigate various model risks. We synthesised three sub-themes and discuss them as follows:

\textbf{Risk mitigation suggestions for downstream developers (108).} Model cards, among other document artefacts, have provided model architectures along with their development process. However, conflicts around development practices between data scientist and software engineers are found, with some software engineers reporting insufficient and hard-to-understand documentation~\cite{nahar2022collaboration}. Therefore, risk mitigation recommendations could help downstream developers bridge the gap between the lack of domain expertise and the necessity for safe model deployment.

The most common recommendation suggests downstream developers to conduct risk assessment tailored to their specific use cases ($\sim29.6\%$). A similar mitigation recommendation involves requiring them to further fine-tune the model for their use cases ($\sim14.8\%$). However, other than a few instances which provide initial resources for reference (e.g.,~\cite{hfdistilbert}), both types of suggestions lack substantive information that developers can follow. For instance, \texttt{stabilityai/StableBeluga-13B} notes, ``\textit{Therefore, before deploying any applications of Beluga, developers should perform safety testing and tuning tailored to their specific applications of the model.}''.

Furthermore, the current model card documents also contain prevalent mitigation suggestions of a more generic nature. These include encouraging downstream developers to familiarise themselves with potential model risks, to inform end users about these risks, as well as abstract references to taking "appropriate" measures to mitigate model risks. These suggestions collectively account for $\sim33.3\%$ of the total recommendations.

In contrast, suggestions on adopting additional techniques to censor the input or output of the model are more concrete and could lead to actionable solutions, accounting for $\sim22.2\%$ of the total recommendations. For instance, \texttt{Salesforce/ctrl} wrote ``\textit{A second recommendation to further screen the input into and output from the model will be implemented through the addition of a check in the CTRL interface to prohibit the insertion into the model of certain negative inputs, which will help control the output that can be generated.}''.

\textbf{Risk mitigation in the model development phase (48).} In addition to recommendations to developers, practitioners also document their endeavours in mitigating various ethics-related risks for the model. Regarding their efforts during model development, the focus revolves around data. Mitigation measures encompass data pre-processing, meticulous selection of data sources, and data annotators. Furthermore, there are also references to the algorithms utilised during the training phase.

\textbf{Risk mitigation in the post-model development phase (13).} Some documents note that the model has undergone safety evaluations, while others acknowledge their lack of efforts in follow-up risk assessment and mitigation. Notably, \texttt{suno/bark} notes ``\textit{To further reduce the chances of unintended use of Bark, we also release a simple classifier to detect Bark-generated audio with high accuracy}''. This supplementary tool aids in detecting model misuse, providing downstream developers with a mitigation solution.

Various ethics-related risk mitigation practices could also provide insights for downstream developers to leverage. However, these factors are often not the focus of model card documents, and discussions tend to be brief. Instead, practitioners' efforts are typically accompanied by discussions of corresponding ethics-related risks associated with the model, effectively serving as explanations for any deficiencies in the models. This ``forward risk mitigation" can be seen as a very preliminary step towards ``thinking through the instances in which unfairness or harm might arise but that are not yet formally addressed or even recognized'', within the framework of what Elish \cite{Elish2019-we} terms the \textit{moral crumple zone}.

\begin{tcolorbox}[left=1pt, top=1pt, right=1pt, bottom=1pt]
    \textbf{Summary of model risk mitigation}: The most frequent sub-theme is recommendations for downstream developers, but many are too abstract or generic to be actionable. Risk mitigation endeavours by model developers can be divided into processes during and after the model development, with discussions typically being brief.
\end{tcolorbox}

\subsection{Model Use Cases}
Instead of providing usage instructions on the model, this theme specifies restrictions on the permissible uses of the model. This is reminiscent of cautions and legal repercussions of medicinal `off-label' use in the context of medical ethics~\cite{Bell438}, or, closer to home, OSS licenses restricting use/deployment if it causes harm~\cite{githubGitHubRaiselyNoHarm}. We synthesised three sub-themes and discuss them as follows:

\textbf{Out-of-scope use cases (87).} This sub-theme specifies scenarios in which the model should not be used. These out-of-scope use cases arise either because the model is not designed to address the situation or because the stakes for using AI are deemed too high.

Among these discussions, the use of the model for critical tasks or systems is most frequent, accounting for $\sim52.9\%$ of the discourse. Critical tasks include generating factually-correct content. For example, some model cards emphasise that "\textit{The model was not trained to provide factual or true representations of people or events, and therefore using the model to generate such content is out-of-scope for its abilities.}" Similar to the model risk of hallucination, this statement emphasises that the model should not engage in such tasks. Tasks involving critical decisions or actions against human benefits are also considered critical. For instance, \texttt{Minej/bert-base-personality} is a model that predicts the personality based on input texts. In the model card document, they note, ``\textit{It should not be used for making critical decisions or judgments about individuals in areas such as employment, education, or legal matters}''. Some instances also mention that the model is not intended for use in safety-critical systems.

Model deployment situations are addressed in $\sim31.0\%$ of the discussions. Some projects state that the direct use of the model without appropriate risk assessment and mitigation is considered out-of-scope. For example, in the project \texttt{tiiuae/falcon-180B}, the statement "\textit{Production use without adequate assessment of risks and mitigation}" is located in the out-of-scope section. Additionally, some projects explicitly mention in their model cards that the model is not intended for production use as they are solely research models. For instance, \texttt{facebook/nllb-200-3.3B} notes, "\textit{NLLB-200 is a research model and is not released for production deployment.}"

Some discussions under this sub-theme also address human interaction with the model, emphasising that the system is not designed for human-facing situations without appropriate guardrails. They also cover unintended use for certain groups and caution against using the model without human supervision.

\textbf{Potential malicious use and misuse of the model (82).} Model malicious use and misuse refer to instances where users are aware of the potential risks and limitations of the model but deliberately use it in ways that cause harm or damage. The most discussed form of misuse is the generation of harmful and offensive content, accounting for $\sim63.4\%$. This is followed by generating disinformation and propaganda ($\sim18.3\%$), violating human privacy ($\sim11.0\%$), and violating copyright or deceiving model behaviour ($\sim6\%$). Notably, the content addressing misuse is typically concise, often presented in bullet points listing considerations, indicating the practitioners have given thoughts to this aspect. 

\textbf{Model term of use (41).} This sub-theme encompasses the licenses, laws, regulations, and policies governing how the model should be used. When stating the licenses, model cards often include corresponding links for readers to follow. However, specific laws and regulations are not typically mentioned in the discussion. For instance, when reading "\textit{Users should also respect the privacy and consent of the data subjects, and adhere to the relevant laws and regulations in their jurisdictions}," in \texttt{KoalaAI/Text-Moderation}, users may not be able to identify which regulations to refer to.

\begin{tcolorbox}[left=1pt, top=1pt, right=1pt, bottom=1pt]
    \textbf{Summary of model use cases}: 
   The out-of-scope sub-theme includes uses in critical tasks, deployment scenarios, and human interaction. Potential malicious use refers to deliberate misuse to cause harm. Model terms of use include licenses, laws, regulations, and policies.

\end{tcolorbox}

\subsection{Reference to other materials}
Containing all information within the document would cause information overload. ``Other materials" refer to other documentation and non-documentation artefacts that are not embedded within the model card document~\cite{gao2023add}. We synthesised two sub-themes and discuss them as follows:

\textbf{Reference to third-party external materials (60). } This sub-theme includes references to external literature, license, responsible usage guidelines, machine learning documents, and external datasets. Many documents refer to additional literature to setup the background for ethics-related issues. ``\textit{Significant research has explored bias and fairness issues with language models (see, e.g., Sheng et al. (2021) and Bender et al. (2021)}'' is a commonly adopted template for laying the background shared among 21 documents. Additionally, materials such as license agreements and responsible usage guidelines provided by large companies are frequently referenced.

\textbf{Reference to in-house materials (34).} Meanwhile, reference are also made to materials developed by the document authors themselves. References are made to blogs, publications, or discussion forums for the corresponding model for additional details. For example, in \texttt{klue/bert-base}, ``\textit{The model developers discuss several ethical considerations related to the model in the paper}'' is presented in the document, with the URL provided to the corresponding paper. For data source of \texttt{HF\_CARD} and \texttt{GH\_README}, documents also make reference to model cards on their GitHub projects. This is because for Hugging Face models, some projects keep their codebase on GitHub, and README files only introduce the repositories.

\begin{tcolorbox}[left=1pt, top=1pt, right=1pt, bottom=1pt]
    \textbf{Summary of reference to other materials}: 
    References deliver additional ethical content beyond the project's scope, providing background context without ``information overload''. These references can point to third-party or in-house materials.
\end{tcolorbox}

\subsection{Others}
There are also codes that do not fit into the previously mentioned themes. They are considered as ``Others''. We synthesised four sub-themes and discuss them as follows:

\textbf{Future socio-ethical research direction (34).} This sub-theme pertains to future directions introduced by this model that have impact on socio-ethical perspectives. The majority code in this sub-theme is related to the potential for the model to be used for future ethical research. Aspirations including developing accessibility tools or calling for better model regulation are also discussed.

\textbf{Limited exploration of ethical considerations (33).} This sub-theme encompasses insufficient efforts on ethical factors. In 21 such Hugging Face model cards, authors used template section structures, and under ethics-related sections, explicit documentation debt such as "\textit{More information needed}" is present. The lack of ethical considerations is also evident in discussions where solving the ethical considerations is deferred to future work, or where users are encouraged to discover and report ethical issues themselves.

\textbf{Model developers' disclaimer (28).} Some documents disclaim responsibility for generated content by the model or consequences from potential model misuse.

\textbf{Environmental impact (15).} Some documents keep track of the energy consumption or carbon emission for training their models; akin to the trend of publishing carbon footprints on flight tickets.


\section{Discussion and Implications}
\subsection{Result Discussion}
We began with an initial set of 3,609 projects from GitHub and Hugging Face. However, after filtering and removing duplicates, only 265 documents were analysed. The small proportion of valid ethics-related documents suggests limited emphasis on documenting ethical considerations in open-source AI models. Additionally, we noted instances of duplicate ethical discussions across documents. It appears that AI model developers did not customise documents for specific cases but rather employed them as templates.

From Table~\ref{tab:qualitative-result}, we observed that the most discussed themes in OSS documents are model behavioural risks, model use cases, and model risk mitigation. Conversely, data quality concerns were covered by the least amount of documents. Meanwhile, only discussion in ``risk mitigation suggestions for downstream developers'' provides solutions to mitigate ethics-related risks, with $\sim28.30\%$ documents covering it. Therefore, the open source AI model documents we collected focus more on articulating ethical problems intrinsic to the model and use case restrictions rather than presenting solutions.


When comparing results across various data sources, \texttt{GH\_README} encompass the least coverage across all themes except for "reference to other materials." This aligns with expectations since README files provide diverse introductory perspectives to a repository, rather than focusing solely on ethical considerations~\cite{prana2019categorizing}. Regarding the disparity between the two model card data sources, \texttt{HF\_CARD} comprises more content in four out of six themes, while \texttt{GH\_CARD} exceeds solely in model use cases and others themes. 

As discussed in the previous section, the theme of ``model behavioural risk'' and the sub-theme of ``risk mitigation suggestions for downstream developers'' contain different levels of detail. During the coding process, we kept side notes of the level of detail being documented. Given the subjective nature of determining the usefulness of a statement, for the model behavioural risk, we categorise the details into ``brief mention'', ``examples'', and ``evaluated result''. Similarly, suggestions were categorised as ``actionable direction'' or  ``abstract advice''. As an illustration, within the examples provided in the results section, the suggestion from \texttt{Salesforcce/ctrl} is categorised as "actionable direction," whereas the suggestion from \texttt{stabilityai/StableBeluga-13B} is deemed ``abstract advice."

Within the subset of documents addressing model behavioural risk, $\sim6.2\%$ offer evaluated results for certain ethical factors, while $\sim18.5\%$ present examples illustrating potential ethics-related risks. The majority of documents ($\sim75.3\%$) merely provide brief mentions of various perspectives on model risks. In terms of risk mitigation recommendations, within the subset of documents addressing this sub-theme, $\sim34.7\%$ offer actionable directions, while the remainder provide abstract advice. Consequently, we identify that most of the ethics-related risks discussed in OSS documents are only briefly outlined. Moreover, the mitigation recommendations provided are limited, and concrete suggestions are even more scarce.

\subsection{Implications}
We provide implications for various stakeholders below.

(1) \textbf{AI Model Developers}. Based on our results, we provide several recommendations to AI model developers when composing ethics-related model card documents. First, they need to allocate more efforts to documenting dataset-related ethical issues, particularly focusing on data curation and its distribution as the dataset is a primary element for model transparency~\cite{pepe2024hugging}. Checklists or tools~\cite{pang2024blip} could help raise awareness of the ethical issues that need to be documented.  We recognise that dedicated documentation artefacts, such as  datasheets~\cite{gebru2021datasheets}, are also used for recording such concerns, but references to such documents are rarely found in our dataset. Second, AI model developers need to provide more details about the identified risks. Due to the expertise gap, some briefly mentioned terminologies might be challenging for documentation users. Providing examples or even concrete evaluations (as illustrated in \cite{mitchell2019model}) can help downstream developers better understand and quantify the issues. Lastly, AI model developers should incorporate more risk mitigation suggestions in their model card documents. While we acknowledge that devising comprehensive suggestions covering all perspectives is challenging, practitioners can take proactive steps to address this issue. They can actively engage with downstream developers to gather their thoughts on risk mitigation and invite successful downstream application collaborators to contribute their solutions, leading to the development of more robust and effective risk mitigation strategies.

(2) \textbf{Software Developers}. First, based on our findings, ethics-related model behavioural risks are inadequately covered. Therefore, downstream software developers should independently conduct comprehensive model risk assessments without solely relying on the provided model cards. Second, since mitigation methods provided are limited, developers can cross-reference documents of models with similar risks, especially those trained on the same dataset or built on the same architecture. Third, developers can contribute to the original model card by adding additionally identified ethical risks and their corresponding mitigation strategies. This collaborative effort can enhance the pre-trained model ecosystem and provide additional mitigation strategies for future users. Last, products exposed to the end-users should prioritise ethical sensitivity. Therefore, we advise software developers to create more concrete ethical documents on top of the model documents that include identified risks, evaluated methods, and mitigation solutions.

(3) \textbf{Researchers}. Researchers can pursue several future directions to enhance ethical documentation practices. First, studying how changes in the models are reflected in ethical documentation could offer insights into documentation maintenance practices and related issues. Second, exploring the alignment between claimed ethical risks in documentation and the actual behaviors of the model would be beneficial. This analysis could identify discrepancies, leading to more accurate documentation. Third, conducting a study from the perspective of downstream developers to understand their views on different ethics-related discussions could offer valuable guidance for improving documentation practices. Understanding the specific needs and preferences of developers could inform the creation of more effective and user-friendly documentation practice.

(4) \textbf{Educators}. Given the insufficient efforts in documenting ethical concerns in current model cards, educators should put more emphasis on ethics~\cite{PaltielCCL23}, and provide practical examples, to ensure that students understand the importance of considering ethics when developing or using AI/ML models. Providing templates and detailed aspects when documenting ethics, such as the findings from our study, could be beneficial for students. This approach would not only raise awareness about ethical considerations in AI models but also equip future practitioners with the knowledge and skills necessary to document ethical concerns effectively in their work.

(5) \textbf{Policy makers}. This work shows clear evidence that the ethical documentation efforts are insufficient even with the use of various proposed documentation frameworks. Therefore, policy makers need to enforce or highlight the specific ethical concerns and required level of detail to be addressed within the documents, prior to model release. In addition, they could also require downstream software developers to conduct more thorough risk assessment and mitigation beyond the points listed in the AI model documents.

\section{Threats to Validity}
This section covers the threats to validity of this study.

\textbf{Internal validity}: We used the GitHub search API to gather relevant GitHub projects containing model cards. We employed "model\_card" and "model-card" as search strings for file names. This approach carries the risk of overlooking model card documents that adhere to different naming conventions. To mitigate this risk, we opted not to perform exact matching. Instead, we allowed for variations in upper and lower case letters, and as long as one of the search strings appeared in the file name, we included the project. We used a keyword-based approach to filter ethics-related documentation. Despite our systematic method of generation and assessment, some documents not explicitly using these terms might have been excluded. Additionally, we did not test software for ethical issues, and documented ethical considerations may not reflect the software's ethical properties. Thus, we advise developers not to rely solely on the provided documentation for ethical assessments.

\textbf{Construct validity}: The subjectivity involved in determining the ethics-related discussions in documents and allocating them into corresponding codes poses a threat to the validity of our results and conclusions. To mitigate this, we rigorously examined the ethics-related content first and resolved any disagreements to reach a consensus via weekly meetings. Regarding the coding themes and results, we cross-referenced with the model card template proposed by Mitchell et al.~\cite{mitchell2019model}. The first author conducted the coding process, with the other two authors closely monitoring how the codes were assigned and synthesised. The coding process and mind map were shared among the group, with consensus achieved. Additionally, to ensure that our analysis is not biased towards statements in duplicate documents, we used cosine distance filter to remove documents that share the same ethical discussions.

\textbf{External validity}: We acknowledge the limitation of the GitHub search API leading to fewer repositories collected from \texttt{GH\_README}. However, since our observation indicates fewer ethical considerations are covered in this source, the impact of the smaller number of documents is minimal. Additionally, our work is limited by the relatively small number of documents. Our results were derived from the analysed repositories, and software documents of other repositories might cover additional ethical considerations. Future work could examine more documents to increase the robustness and generalisability of the findings.  Moreover, considering that different processes are used for proprietary AI models and products, our findings are only limited to open source models and software. To increase the generalisability of our results, we analysed documentation artefacts on both GitHub and Hugging Face. We analysed both README files and model cards, as these data sources are mostly used in OSS for documentation, and therefore, more possible for us to find ethical documentation. Regarding data representativeness, our results are more representative of recent popular AI models due to our sampling strategy. Therefore, our findings might not be indicative of the documentation practices for older and less popular models. However, as a previous study identified~\cite{pepe2024hugging}, less popular repositories tend to document less, so our results should cover comprehensive aspects of ethical documentation practices.

\section{Conclusions}
We investigated the documentation practice of ethical considerations in open-source AI model documents. We developed a keyword set to capture these discussions within the documents, including model cards and READMEs. Using this keyword set to filter on an initially collected data of 2,346 documents, we analysed 265 open source AI model documents, to understand how ethical considerations are being documented in practice. Our thematic analysis yields 81 codes related to six themes: (1) Data quality concerns, (2) Model behavioural risks, (3) Model risk mitigation, (4) Model use cases, (5) Reference to other materials, and (6) Others. 

Our findings reveal that OSS ethical documents primarily focus on articulating ethical problem statements and use case restrictions, rather than presenting solutions. In addition, most ethics-related risk discussions are only briefly outlined, and the already scarce risk mitigation suggestions are mostly not concrete and actionable. In our discussion, we provide suggestions for different stakeholders to enhance documentation practice in ethical considerations.

\section{Acknowledgements}
This work was supported by seed funding for enhancing research collaboration within the CS group at the School of Computing and Information Systems (CIS), University of Melbourne.

\bibliographystyle{ACM-Reference-Format}
\bibliography{reference}

\end{document}